\documentclass[twocolumn,showpacs,amsmath,amssymb]{revtex4}
\usepackage{graphicx}
\usepackage{dcolumn}
\usepackage{bm}
\usepackage[dvips]{epsfig}

\newcommand{\ba}{\begin{eqnarray}}
\newcommand{\ea}{\end{eqnarray}}
\newcommand{\ban}{\begin{eqnarray*}}
\newcommand{\ean}{\end{eqnarray*}}
\newcommand{\be}{\begin{equation}}
\newcommand{\ee}{\end{equation}}
\newcommand{\bd}{\begin{displaymath}}
\newcommand{\ed}{\end{displaymath}}

\newcommand{\n}[1]{\label{#1}}

\newcommand{\eq}[1]{(\ref{#1})}

\newcommand{\hh}{\, ,\hspace{0.5cm}}

\begin{document}

\title{Hidden Symmetry of Higher Dimensional Kerr--NUT--AdS Spacetimes}
\author{David Kubiz\v n\'ak and Valeri P. Frolov}
\affiliation{Theoretical Physics Institute, University of Alberta,
Edmonton, Alberta, Canada, T6G 2G7}

\email{kubiznak@phys.ualberta.ca}
\email{frolov@phys.ualberta.ca}

\date{\today}

\begin{abstract} 
It is well known that $4$--dimensional Kerr--NUT--AdS spacetime
possesses the hidden symmetry associated with the Killing--Yano tensor.
This tensor is `universal' in the sense that there exist  coordinates
where it does not  depend on any of the free parameters of the metric. Recently the
general higher dimensional Kerr--NUT--AdS solutions of the Einstein
equations were obtained. We demonstrate that all these metrics  with
arbitrary rotation and NUT parameters admit a universal Killing--Yano
tensor. We give an explicit presentation of the Killing--Yano tensor and 
associated second rank Killing
tensor and briefly discuss their properties.
\end{abstract}

\pacs{04.70.Bw, 04.50.+h, 04.20.Jb \hfill  Alberta-Thy-10-06}

\maketitle

There are several reasons why higher dimensional black hole solutions
attracted a lot of attention recently. The string theory is consistent
only when the number of spacetime dimensions is either $10$ or $26$. Black
holes in the string theory were widely discussed in connection with the
problem of microscopical explanation of the black hole entropy. Also in the
recent models with large extra dimensions it is assumed that one or more
additional spatial dimensions are present. In such models one expects mini
black hole production in the high energy collisions of particles.  Mini
black holes can serve as a probe of the extra dimensions. At the same time their
interaction with the brane, representing our physical world, can give the 
information about the brane properties. 

Higher dimensional non--rotating black hole solutions were found long
time ago by Tangherlini \cite{tang}. The solutions for  rotating black
holes, which are higher dimensional generalization of the Kerr metric,
were obtained by Myers and Perry (MP metrics) \cite{MP}. More recently
the MP solutions were generalized to include the cosmological
constant \cite{Hawk, Page1, Page2}. These solutions are of special interest in
connection with their possible applications for the study of AdS/CFT
correspondence. Further generalization of the higher dimensional
Kerr--AdS solutions which includes also the NUT parameters was found in
\cite{Pope, Pope2}.

The higher dimensional Kerr--NUT--AdS metrics are stationary and
axisymmetric; they possess the Killing vectors which generate
the time translation and rotations in the independent $2$D planes of rotation.
In this paper we show that besides these evident symmetries all higher
dimensional Kerr--NUT--AdS metrics, describing the rotating black holes
with arbitrary rotation and NUT parameters in an
asymptotically AdS spacetime, have a new hidden symmetry.

Hidden symmetries of $4$D black hole solutions of the Einstein equations
are well known. The study of them has begun when 
Carter \cite{cart} discovered that the Hamilton--Jacobi 
equation in the (charged) Kerr metric allows the separation of variables.
Walker and Penrose \cite{WP} demonstrated that this separability is a consequence 
of the existence of an irreducible  second rank Killing tensor 
\be\label{KT}
K_{\mu\nu}=K_{(\mu\nu)}\hh K_{(\mu\nu;\lambda)}=0\, .
\ee 
Penrose and Floyd \cite{Penrose} found that this Killing tensor
is a `square' of a more fundamental antisymmetric Killing--Yano
(KY) tensor  \cite{Yano} 
\be
f_{\mu\nu}=f_{[\mu\nu]}\hh
f_{\mu(\nu;\lambda)}=0\, .
\ee
The second rank Killing tensor can be expressed in terms of $f_{\mu\nu}$
as follows
\be
K_{\mu\nu}=f_{\mu\alpha}f^{\ \alpha}_{\nu}\, .\n{square}
\ee
Carter \cite{Carter} showed that the Killing--Yano tensor itself is derivable from the 
existence of a `Killing--Maxwell' form (an analogue of what we call below a potential $b$).

The existence of the Killing--Yano tensor for the Kerr metric is a
consequence of the fact that this metric belongs to the Petrov type D.
Collinson \cite{Coll74} proved that if a vacuum solution of 4D Einstein equations  
admits a  KY tensor it belongs to the type D. All the vacuum type D
solutions were obtained by Kinnersley \cite{Kinn}. Demia\'nsky and
Francaviglia \cite{DeFr} showed that in the absence of acceleration
these solutions admit the KY tensor. 

The type D solutions of the Einstein--Maxwell equations with the
cosmological constant allow a convenient representation in the form
of the Pleba\'{n}ski--Demia\'{n}ski metric \cite{PlDe} (for a recent
review and  reinterpretation of parameters in this solution see \cite{GrPo}).  A
subclass of solutions without acceleration studied by Pleba\'{n}ski
\cite{Pleb75} possesses a Killing--Yano tensor. The Pleba\'{n}ski
metric reads
\be\label{Pleb}
ds^2_{4}=Q_p(d\tau-r^2d\sigma)^2-Q_r(d\tau+p^2d\sigma)^2+
\frac{dr^2}{Q_r}+\frac{dp^2}{Q_p}\, ,
\ee
where, in the absence of electric and magnetic charges, 
\begin{equation}\label{QrQp}
Q_p=\frac{\gamma+2lp-\epsilon p^2-\lambda p^4}{r^2+p^2},\,
Q_r=\frac{\gamma-2mr+\epsilon r^2-\lambda r^4}{r^2+p^2}\,.
\end{equation}
This metric obeys the equation $R_{\mu\nu}=3\lambda g_{\mu\nu}$.
Its form is invariant under the rescaling of the coordinates $p\to \alpha p$, 
$r\to \alpha r$, $\tau\to \alpha^{-1}\tau$, $\sigma\to \alpha^{-3}\sigma$.
Under this transformation the cosmological constant parameter 
$\lambda$ is invariant, while the other parameters
change. One can always use this transformation to fix the magnitude
of one of the parameters, say $\epsilon$. Afterwards the 
parameters $(m, \gamma, l)$ are related to the mass, angular momentum
and NUT charge (see. e.g., \cite{Pope}). The Killing--Yano tensor (which
under the scaling transformation is multiplied by a constant) reads
\begin{equation}\label{yanoPleb}
f^{(4)}=rdp\!\wedge\!(d\tau-r^2d\sigma)+pdr\!\wedge\!(d\tau+p^2d\sigma).
\end{equation} 
It is interesting that in the chosen coordinates 
its form
does not depend on the parameters $(\lambda,\gamma,m,l)$. Moreover,
one can easily check, using GRTensor, that $f^{(4)}$ remains the KY
tensor for the solutions of the cosmological Einstein--Maxwell
equations \eq{Pleb} when the electric and magnetic charges are
included in (\ref{QrQp}). Let us emphasize that this {\em universality
property} is valid only for a  specially chosen coordinate system, 
but the very existence of such coordinates is rather non--trivial.

The hidden symmetries of higher dimensional rotating black holes were
first discovered for $5$D MP metrics in \cite{FSa,FSb}. Namely, it was
demonstrated that both, the Hamilton--Jacobi and massless scalar field
equations, allow the separation of variables; the corresponding
Killing tensor was obtained. This result was generalized for the 
MP metrics in an arbitrary number of dimensions, provided that their rotation
parameters can be divided into two classes, and within each of the
classes the rotation  parameters are equal one to another \cite{mur1}.
A similar result is valid in the presence of the cosmological constant
\cite{mur2,mur3} and NUT parameters \cite{Pope, Pope3, davis}. Recently we
explicitly demonstrated that the KY and Killing tensors exist for any MP metric with
arbitrary rotation parameters \cite{FrKu1}. We generalize now this result to
the Kerr--NUT--AdS spacetimes. 

Our starting point is the expression for the general higher dimensional
Kerr--NUT--AdS metrics obtained recently \cite{Pope2}. For notation
convenience, we deal with an analytical continuation of these metrics.
Let $D$ denotes the total number of spacetime dimensions. We
define $n=[D/2]$ and for brevity $\varepsilon=D-2n$,
$m=n-1+\varepsilon$. The metrics read
\begin{equation}\label{metrics}
ds^2=\sum_{\mu=1}^n[\frac{dx_{\mu}^2}{Q_{\mu}}+Q_{\mu}(\sum_{k=0}^{n-1}A_{\mu}^{(k)}d\psi_k)^2]
-\frac{\varepsilon c}{A^{(n)}}(\sum_{k=0}^nA^{(k)}d\psi_k)^2\!\!,
\end{equation}
where 
\begin{eqnarray}\label{co}
Q_{\mu}&=&\frac{X_{\mu}}{U_{\mu}},\quad 
U_{\mu}=\prod_{\nu=1}^{\prime\,n}(x_{\nu}^2-x_{\mu}^2),\quad c=\prod_{k=1}^m a_k^2, \nonumber\\ 
X_{\mu}&=&(-1)^{\varepsilon}\frac{g^2x_{\mu}^2-1}{x_{\mu}^{2\varepsilon}}\prod_{k=1}^{m}(a_k^2-x_{\mu}^2)
+2M_{\mu}(-x_{\mu})^{1-\varepsilon},\nonumber\\
A_{\mu}^{(k)}&=&\!\!\!\!\!\sum_{\nu_1<\dots<\nu_k}^{\prime}\!\!\!\!\!x^2_{\nu_1}\dots x^2_{\nu_k},\ 
A^{(k)}=\!\!\!\!\!\sum_{\nu_1<\dots<\nu_k}\!\!\!\!\!x^2_{\nu_1}\dots x^2_{\nu_k}.
\end{eqnarray}
The primes on the sum  and product symbols indicate that the
index $\nu=\mu$ is omitted. In odd dimensions the apparent singularity
for $c=0$ is removed when $c$ is `absorbed' in the definition of
$\psi_n$. 

The physical metrics with proper signature are recovered when a standard
radial coordinate $r=-ix_n$ is introduced.  The metrics possess
$(n+\varepsilon)$ Killing vectors $\partial_{\psi_k}$. $\Psi_0$ plays
the role of the time coordinate. The meaning of the parameters is the
following: $a_k$ denote $m$ `rotation' parameters, $M_{\alpha}$ for
$\alpha=1\dots (n-1)$ denote the `NUT' parameters,
$M=-i^{1+\varepsilon}M_n$ is the mass and $\lambda=-g^2$ is proportional to 
the cosmological constant \cite{fn}. In odd dimensions one of the `NUT' parameters may
be eliminated due to the scaling symmetry (see \cite{Pope2} for details).
Therefore these metrics constitute $(D-1-\varepsilon)$-parametric
solutions of the cosmological Einstein equations
$R_{\mu\nu}=(D-1)\lambda g_{\mu\nu}$. Let us finally remark that the
formulas (\ref{metrics}) and (\ref{co}) are applicable also in $D=3$
where  one recovers $2$-parametric BTZ black hole \cite{BTZ}. 

The connection with the Kerr--AdS metrics \cite{Page1,Page2} is established through the `Jacobi' transformation of 
the (constrained) `latitude' Boyer--Lindquist  coordinates $(i=1\dots
n, a_{m+1}=0)$
\begin{equation}
\mu_i^2={\normalsize 
\prod_{\alpha=1}^{n-1}(a_i^2-x_{\alpha}^2)/\!\!\!\!\! 
\prod_{k=1,\ k\neq i}^{n}\!\!\!\!(a_i^2-a_k^2)},
\end{equation}
and the due rescaling of $\psi_k$ coordinates.

The inverse metrics read 
\begin{eqnarray}\label{inverse}
(\partial_s)^2&=&
\sum_{\mu=1}^n\frac{x^{-4\varepsilon}_{\mu}}{Q_{\mu}U_{\mu}^2}
\,[\sum_{k=0}^{m}(-1)^kx_{\mu}^{2(m-k)}\partial_{\psi_{k}}]^2\nonumber\\
&+&\sum_{\mu=1}^nQ_{\mu}(\partial_{x_{\mu}})^2
-\frac{\varepsilon}{cA^{(n)}}(\partial_{\psi_{n}})^2.
\end{eqnarray}

Our claim is  that the metrics (\ref{metrics}) admit the 
$(D-2)$--rank Killing--Yano tensor and the second rank Killing tensor.
In a general case the KY tensor  is defined as a $p$--form 
$f$ which obeys the equation
\be
\label{Yeq}
\nabla_{\!(\alpha_1}f_{\alpha_2)\alpha_3\dots \,\alpha_{p+1}}=0\, .
\ee 
The associated second rank Killing tensor is 
\begin{equation}\label{square}
K_{\mu\nu}=\frac{1}{(p-1)!}\,f_{\mu\alpha_1\dots\,\alpha_{p-1}}
f_{\nu}^{\ \alpha_1\dots \alpha_{p-1}}\, . 
\end{equation}

We shall follow the procedure established in \cite{FrKu1}.
Instead of dealing with the KY tensor of the rank $(D-2)$, it is technically
easier to consider its dual tensor $k_{\mu\nu}$ related
to $f$ as
 \begin{equation}\n{fk}
f_{\alpha_1\dots \alpha_{D-2}}=(*k)_{\alpha_1\,\ldots \mu_{D-2}}
={1\over 2}\,e_{\alpha_1\,\ldots
\alpha_{D-2}\kappa\delta}k^{\kappa\delta}\, .
\end{equation}
Here  $e_{\alpha_1\! \ldots \alpha_D}$ is the totally  antisymmetric tensor
\be
e_{\alpha_1\!\ldots \alpha_D}=
\sqrt{-{\rm g}}\, \epsilon_{\alpha_1\!\ldots \alpha_D}\, .
\ee
It is possible to show that $k$ is a conformal Killing--Yano (CYK) tensor \cite{Tachibana, Kaschiwada}, obeying the
equation
\begin{equation}\n{cye}
k_{\alpha\beta;\,\gamma}+k_{\gamma\beta;\,\alpha}
=\frac{2}{D-1} (g_{\alpha\gamma}\,k^{\sigma}_{\
\beta;\,\sigma}+ g_{\beta(\alpha}k_{\gamma)\, \, \,  ;\,\sigma}^{\,
\, \,  \,\sigma} )\,.
\end{equation}
If the second rank CYK tensor $k$ is closed, $dk=0$, the form $f$ defined by \eq{fk} is the
KY tensor \cite{Cari}. In the latter case the form $k$ can be written
(at least locally) as $k=db$, where $b$ is a one--form (potential).

We propose the following
ansatz for potential $b$
\begin{equation}\label{cykp}
2b=\sum_{k=0}^{n-1}A^{(k+1)}d\psi_k.
\end{equation}
Its exterior derivative 
\ba\n{cyk}
2k=2db=\sum_{k=0}^{n-1}dA^{(k+1)}\!\wedge d\psi_k
\ea
is the $2$--rank CKY tensor. We have
explicitly checked, using GRTensor, that the equations \eq{cye} are satisfied
for $D\leq 9$. We strongly believe that they are valid in any number of
dimensions.

The quantities $A^{(k)}$ and $A_{\mu}^{(k)}$ contain no
dependence on the parameters specifying the solution \eq{metrics}.
The special structure of $k_{\alpha\beta}$   together with the form of
inverse metrics (\ref{inverse}) imply that $k^{\alpha\beta}$ has the same property.
The determinant of (\ref{metrics}) is
\begin{equation}\label{det}
{\rm g}=(-cA^{(n)})^{\varepsilon}[{\rm det}A_{\mu}^{(k)}]^2.
\end{equation}
In this expression $A_{\mu}^{(k)}$ is understood as the $n\times
n$ matrix. The relation (\ref{fk}) implies that the  KY tensor $f_{\mu_1\ldots \mu_{D-2}}$
has no dependence on the parameters (up to a not essential common constant factor $\sqrt{c}$ in
odd dimensions). In other words, $f$  is universal.

Using (\ref{inverse}), \eq{fk}, \eq{cyk}  and (\ref{det}) one can easily
obtain KY tensor $f$ in an explicit form for an arbitrary number of dimensions.
Since $D=3$ case is trivial ($f$ is a Killing vector) we shall give the
expressions in $D=4, 5$ and $6$. 

In $D=4$ we denote $x_1=p,\ x_2=ir,\ \psi_0=\tau,\ \psi_1=\sigma$. Then we recover
the KY tensor (\ref{yanoPleb}). 

In $D=5$ we find
\begin{equation}
f^{(5)}=x_1dx_1\!\wedge h(x_2)+x_2dx_2\!\wedge h(x_1),
\end{equation}
where we have defined  
\begin{equation}
h(x)=d\psi_0\!\wedge\! d\psi_1+x^2d\psi_0\!\wedge\! d\psi_2+x^4d\psi_1\!\wedge\! d\psi_2.
\end{equation}

Finally in $D=6$ let us denote $x_1=x,\ x_2=y,\ x_3=z$. Then we have
\begin{eqnarray}
f^{(6)}&\!\!=\!\!&z(x^2-y^2)dx\!\wedge\!dy\!\wedge h(z)+y(x^2-z^2)dx\!\wedge\!dz\!\wedge h(y)\nonumber\\
&\!\!+\!\!&x(y^2-z^2) dy\!\wedge\!dz\!\wedge h(x).
\end{eqnarray}

The existence of a KY tensor immediately implies the existence of a Killing tensor.
With the help of CKY one can rewrite the formula (\ref{square}) as 
\be\label{jak}
K_{\mu\nu}=Q_{\mu\nu}-{1\over 2}\, g_{\mu\nu}Q^{\alpha}_{\ \alpha},
\ee
where $Q_{\mu\nu}=k_{\mu\alpha}k_{\nu}^{\ \alpha}$ is the conformal Killing tensor.
The components of (conformal) Killing tensor are most easily written in `mixed' indices. We find
\begin{equation}
Q^{\alpha}_{\ \beta}=\sum_{\mu=1}^n\delta^{\alpha}_{\,x_{\mu}}\delta^{x_{\mu}}_{\,\beta}x_{\mu}^2+
\sum_{k=0}^{n-1}\delta^{\alpha}_{\,\psi_0}\delta^{\psi_k}_{\,\beta}A^{(k+1)}-
\sum_{k=0}^{m-1}\delta^{\alpha}_{\,\psi_{k+1}}\delta^{\psi_k}_{\,\beta}.
\end{equation}
Evidently, $Q^{\alpha}_{\ \alpha}=2A^{(1)}$, and so we have
\begin{equation}
K^{\alpha}_{\ \beta}=Q^{\alpha}_{\ \beta}-A^{(1)}\delta^{\alpha}_{\,\beta}.
\end{equation}
In this `mixed' form the (conformal) Killing tensor is universal. With both indices down, $K_{\mu\nu}$
possesses the same structure of diagonal and nondiagonal sectors as the metric.

Similar to the Myers--Perry case the constructed KY and Killing tensors have direct connection with
the isometries of the spacetime. We find
\be
\xi^{\mu}={1\over D-1}\,k^{\sigma\mu}_{\ \ \ ;\,\sigma}=(\partial_{\psi_0})^{\mu},
\ee
and also
\begin{equation}
\eta^{\mu}=-K^{\mu}_{\ \alpha} \xi^{\alpha}=(\partial_{\psi_{\!1}})^{\mu}.
\end{equation}

To summarize, we claim that all the general Kerr--NUT--AdS metrics
\eq{metrics} in any number of dimensions and with arbitrary 
parameters possess both the Killing--Yano and Killing tensors. The
existence of an additional integral of motion is implied. However,
for $D\ge 6$ the total number of integrals of motion, connected with
the Killing vectors and one additional Killing tensor, is not
sufficient for the separation of variables in the Hamilton--Jacobi,
Dirac, and Klein--Gordon equations. An interesting open question is
whether there exist other non--trivial Killing--Yano and Killing
tensors in the spacetime \eq{metrics}. It would be also interesting
to understand whether in  higher dimensions there exists a deep
relation between the hidden symmetries and the algebraic type of the
metric, similar to what has been observed in the four dimensional
case.

\noindent  

\section*{Acknowledgments}  
\noindent  
One of the authors (VF) thanks the Natural Sciences and Engineering
Research Council of Canada and the Killam Trust for the financial
support. He also thanks the Physics Department of the University of
Tours for the hospitality during the work on the paper. The other author
(DK) is grateful to the Golden Bell Jar Graduate Scholarship in
Physics at the University of Alberta.



\begin{thebibliography}{}
\bibitem{fn} We use notations `rotation' and `NUT' parameters to stress
that the meaning of these parameters depends on the choice of the
scaling  fixing condition.
\end{thebibliography}
\end{document}